# Frequency-tunable metamaterials using broadside-coupled split ring resonators

**Evren Ekmekci[1, 3, 4], Andrew C. Strikwerda[1], Kebin Fan[2], Xin Zhang[2], Gonul Turhan-Sayan[3], and Richard D. Averitt[1]**

[1]Boston University, Department of Physics, Boston, MA 02215, USA
[2]Boston University, Department of Mechanical Engineering, Boston, MA 02215, USA
[3]Middle East Technical University, Department of Electrical and Electronics Engineering, 06531, Ankara, Turkey
[4]Suleyman Demirel University, Department of Electronics and Communication Engineering, 32260, Isparta, Turkey

**We present frequency tunable metamaterial designs at terahertz (THz) frequencies using broadside-coupled split ring resonator (BC-SRR) arrays. Frequency tuning, arising from changes in near field coupling, is obtained by in-plane horizontal or vertical displacements of the two SRR layers. For electrical excitation, the resonance frequency continuously *redshifts* as a function of displacement. The maximum frequency shift occurs for displacement of half a unit cell, with vertical displacement resulting in a shift of 663 GHz (51% of $f_0$) and horizontal displacement yielding a shift of 270 GHz (20% of $f_0$). We also discuss the significant differences in tuning that arise for electrical excitation in comparison to magnetic excitation of BC-SRRs.**

Across the spectrum, electromagnetic metamaterials research continues to blossom with a significant fraction of work focusing on split ring resonators (SRRs) [1] and their variants which includes, as examples, spiral [1], labyrinth [2], and electric field coupled (ELC) resonators [3]. It is well known that metamaterials can provide a $\mu$-negative (MNG) and/or $\varepsilon$-negative (ENG) response. However, these operational MNG/ENG bands are limited to a narrow spectral region. Frequency tunable metamaterials have become an important area of interest since tuning the resonance frequency ($f_0$) can effectively extend the operating bandwidth for certain applications.

Numerous approaches to modify the electromagnetic response of metamaterials (i.e. frequency and/or amplitude of the resonance) have been considered. This includes modification of the substrate parameters (e.g. permittivity) [4, 5], the use of liquid crystals [6–7], lumped capacitors or varactors [8-12], ferromagnetic [13, 14] and ferroelectric [15, 16] techniques, semiconductors [17-19], microelectromechanical (MEMS) switches [20-22], and tuning based on near-field interactions between adjacent SRRs [23-27]. This last approach is the topic of the present study.

Wang *et al.* proposed tuning the electromagnetic response in the microwave region through sub-unit cell relative displacements between the two layers comprising a broadside-coupled SRR (BC-SRR) structure (see Fig. 1(b)) under magnetic excitation [23]. They showed that displacement resulted in a capacitance change (the mutual inductance change was negligible) yielding a shift of $f_0$. In [24] and [25] structural tuning was also investigated under magnetic excitation at microwave frequencies by in-plane vertical displacement of the BC-SRR layers. The definition of in-plane horizontal and vertical displacement of the two layers comprising the BC-SRR array is depicted in Fig. 1(c) and (d). Importantly, as a function of horizontal displacement, a blueshift was observed for magnetic excitation [25].

In this paper, we investigate the effects of lateral shifting of broadside coupled metamaterials at THz frequencies under electrical excitation (see Fig 1(b)). As we demonstrate, the near field coupling of these structures is different for electrical and magnetic excitation. To illustrate this, we design, fabricate, and characterize two-layered terahertz (THz) SRR arrays with different lateral shifts between the two layers. Arrays with vertically and horizontally displaced BC-SRRs are characterized experimentally using THz time-domain spectroscopy (THz-TDS) and modeled using CST Microwave Studio. Of greatest significance is that for electrical excitation, the resonance frequency continuously *redshifts* as a function of displacement in contradistinction to magnetic excitation. The maximum frequency shift occurs for displacement of half a unit cell, with vertical displacement resulting in a shift of 663 GHz (51% of $f_0$) and horizontal displacement yielding a shift of 270 GHz (20% of $f_0$) . We present an intuitive description of changes in coupling in BC-SRRs as a function of displacement to provide insight into the different electromagnetic responses under electrical and magnetic excitation.

To gain insight into the electromagnetic response of shifted BC-SRR structures, we qualitatively examine the current and charge distributions of the top and bottom layers of a single unit cell under electrical and magnetic excitation. In the following, we consider exclusively the LC resonance where $f_0 \sim (L_{total}\, C_{total})^{-1/2}$ (i.e. not the higher frequency dipolar resonances). Fig. 2(a) shows that under magnetic excitation, the surface currents of the top and bottom layers are in the same direction. This results in a mutual inductance ($L_{mut}$) with a positive sign yielding a total inductance $L_{total} = L_{self} + L_{mut}$. For electrical excitation (Fig. 2(b)), the surface currents are in opposite directions meaning that $L_{mut}$ is negative resulting in a total inductance $L_{total} = L_{self} - L_{mut}$.. Thus, lateral shifts between the top and bottom layers in either the horizontal or vertical direction will decrease/increase $L_{total}$ under magnetic/electrical excitation. Similarly, the surface charge distribution and subsequent mutual capacitance depends on the excitation conditions. For magnetic excitation, positive and negative charges between the two layers overlap resulting in a significant mutual capacitance. While this description is qualitative, it shows that, under magnetic excitation, lateral shifting (either horizontal or vertical) decreases the inductance and capacitance consistent with the blueshift observed in the previous work [25].

In contrast, for electrical excitation, negative charges overlap with negative and positive with positive between the two layers yielding a much weaker or negligible mutual capacitance. Unlike the mutual inductance, the shifting direction becomes important. For vertical shifting, the positive and negative charged arms between the upper and lower layers of adjacent unit cells come closer (i.e. total overlap for a shift of ½ a unit cell) leading to an increased mutual capacitance. However, for horizontal shifting, the SRR arrays slide over the SRR arms with charges of the same sign providing a smaller change in the mutual capacitance. This suggests a greater frequency shift for vertical displacement in comparison to horizontal displacement under electrical excitation. In addition, it is clear that for electrical excitation, lateral displacement should lead to redshifting of the LC resonance since both the capacitance and inductance increase.

To measure changes in the resonant response as a function of lateral displacement under electrical excitation, we fabricated structures composed of square shaped SRR unit-cells having the same physical dimensions as shown schematically in Fig. 3. The unit-cell periodicity is $P$=58µm, metallization side-length $l$=40µm, metallization width $w$=11µm, gap width $g$=5µm. This choice of dimensions means that a 30µm shift will create a perfect overlap between legs of a top layer SRR with the legs of two bottom layer SRRs in adjacent unit cells.

The structures were fabricated using conventional photolithography. For all structures, 5µm of polyimide was spin-coated on GaAs as a superstrate and then 200nm thick gold with a 10nm thick adhesion layer of titanium is deposited on a resist layer (S1813, Shipley) and patterned to form a planar array of SRR structures. Another 4µm thick polyimide layer was then coated on the SRR array as the spacer. Next, the second planar array of SRRs was patterned to form a BC-SRR structure. Finally, a 5µm thick polyimide was coated on metamaterials as the second superstrate (see Fig. 3(b)). The multi-layer structure was removed from the GaAs resulting in a thin metamaterial film with a total thickness of 14 µm. The polyimide (PI-5878G, HD MicrosystemsTM) has the relative permittivity ($\varepsilon_r$) of 2.88 and the dielectric loss-tangent of ($\tan\delta_c$=0.0313). The measurements were performed using THz-TDS.

To characterize changes in the electromagnetic response resulting from changes in near field coupling, we fabricated, tested, and simulated structures with shifts from 0μm to 30μm in 5μm steps. Optical microscope pictures of the vertically shifted BC-SRR arrays and the corresponding simulation/experimental results are given in Fig. 4(a) and Fig. 4(b)/Fig. 4(c), respectively. With increasing displacement, the resonance frequency decreases dramatically from 1.351 THz to 0.631 THz in simulation and, experimentally, from 1.304 THz to 0.641 THz. These results correspond to a 720 GHz absolute and 53% percentage shift in simulation and a 663 GHz absolute and 51% percentage shift in experiment.

Figure 5(a) and 5(b)/(c) show optical microscope pictures of horizontally displaced BC-SRR arrays and their corresponding simulation/experimental results. As the shift increases from 5μm to 30μm, the resonance decreases from 1.376 THz to 1.100 THz in simulation and from 1.328 THz to 1.058 THz in experiment. These results correspond to a 276 GHz absolute and 20% percentage shift in simulation and 270 GHz absolute and 20% percentage shift in experiment. It should be noted that this reduced shift (in comparison to vertical displacement) is anticipated from our basic coupling discussion above.

Finally, we note that as a function of shifting (especially obvious for vertically shifting), there is a broadening of the transmission resonance. The analysis of this broadening (arising from an increase in the oscillator strength), along with an effective medium analysis will be presented in a subsequent publication.

In conclusion, we have investigated the effect of near field coupling in electrically excited BC-SRR arrays. We have demonstrated two tunable metamaterial structures, including one which, experimentally, shifts its resonance frequency by over 50% of the center frequency. Further, we have highlighted the substantial differences in the electromagnetic response that arise for electrical excitation in comparison to magnetic excitation. Our results demonstrate the use of structural tunability in THz region under electrical excitation, and highlight the significance of near field coupling of SRRs for future metamaterial applications.

Evren Ekmekci acknowledges TUBITAK for supporting part of his Ph.D. studies through Program 2214 while he is in Boston University. We acknowledge partial support from DOD/Army Research Laboratory under Contract No. W911NF-06-2-0040, AFOSR under Contract No. FA9550-09-1-0708, NSF under Contract No. ECCS 0802036, and DARPA under Contract No. HR0011-08-1-0044 (HT, ACS, ML, EE, KF, XZ and RDA). The authors would also like to thank the Photonics Center at Boston University for all of the technical support throughout the course of this research.

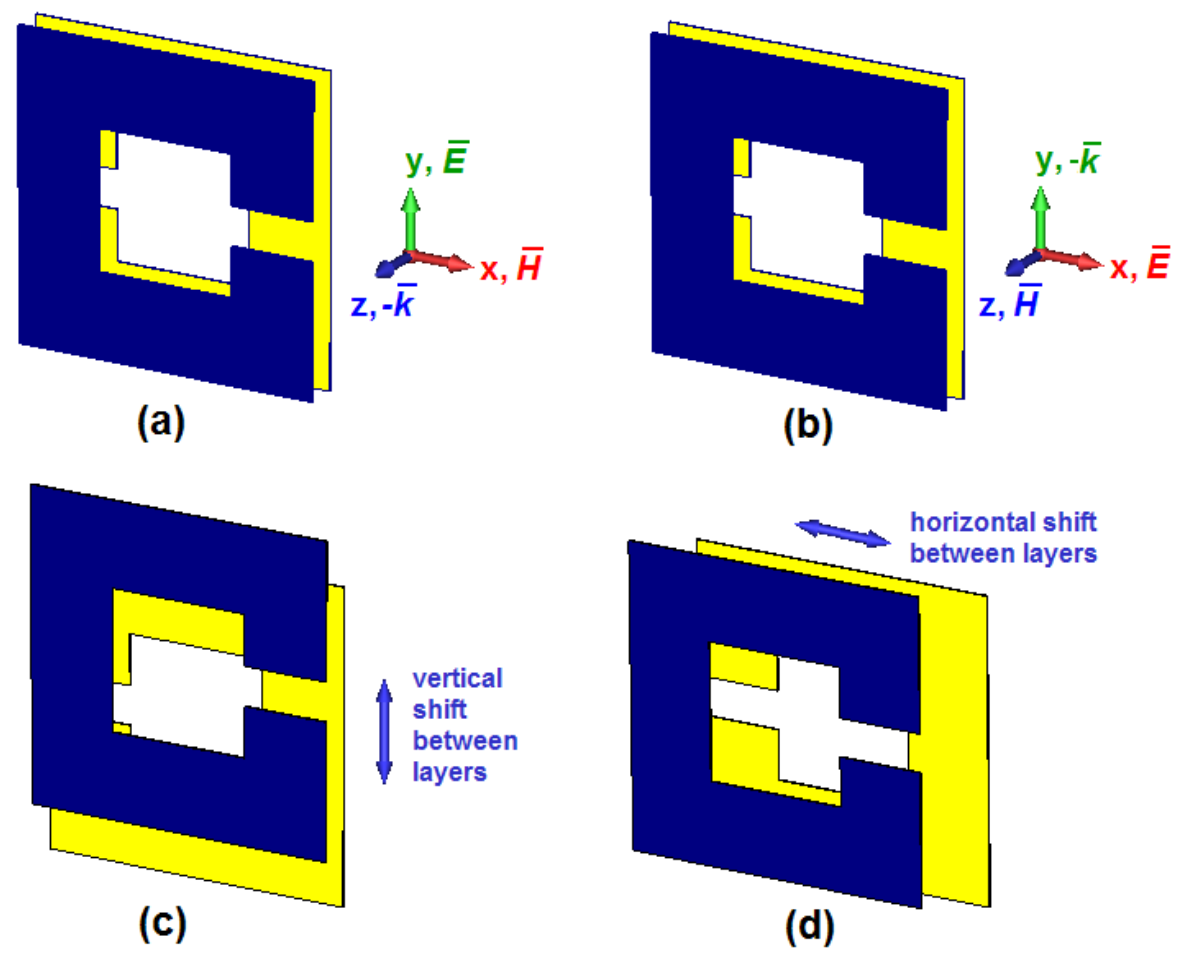

Fig. 1 . BC-SRR under (a) electrical and (b) magnetic excitation. (c) Vertically and (d) horizontally shifting between two layers.

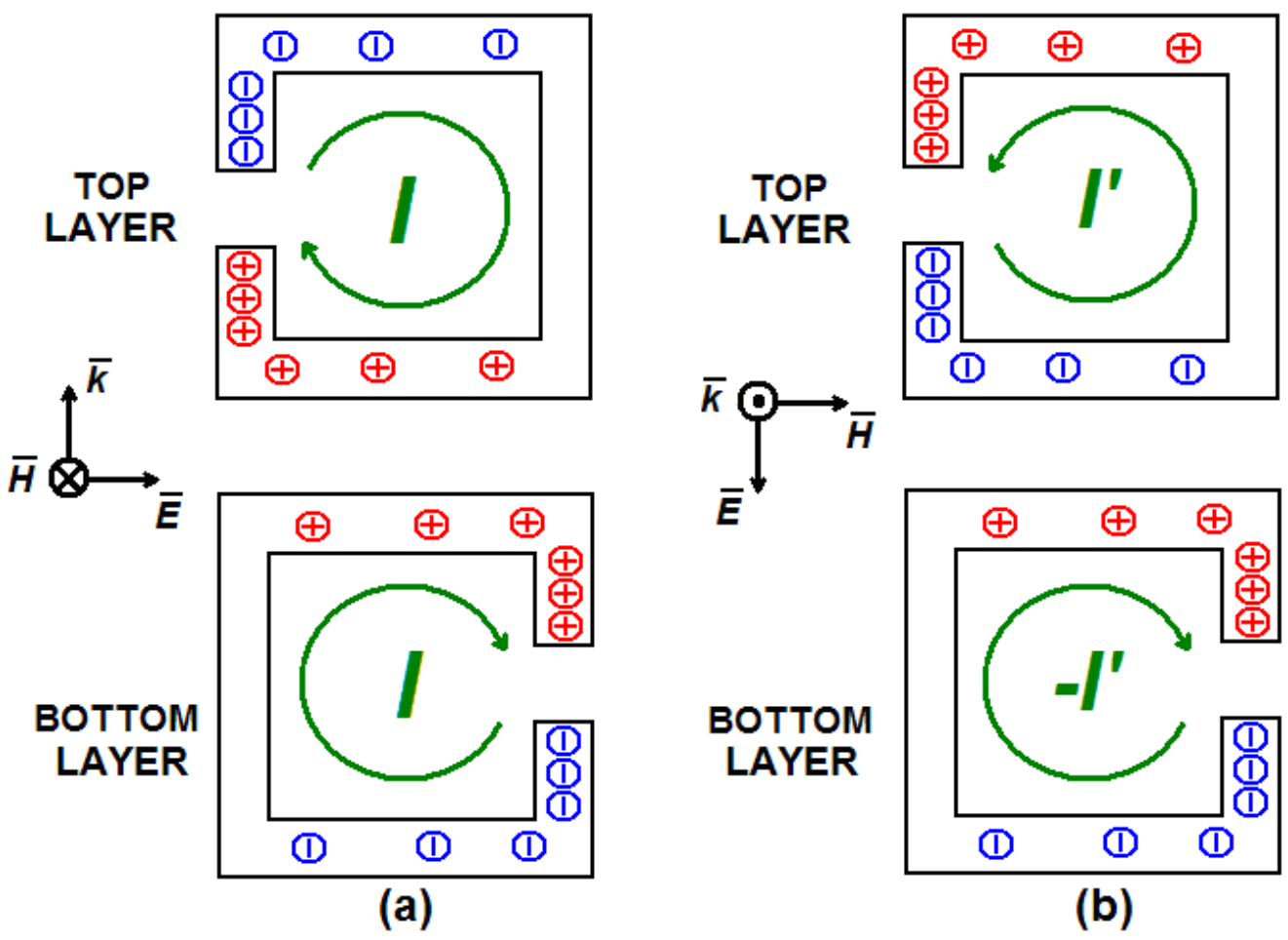

Fig. 2. Schematic views of the surface charge distributions and their corresponding surface current distributions for (a) magnetically and (b) electrically excited BC-SRR structures. The top and bottom layers are shown displaced for clarity.

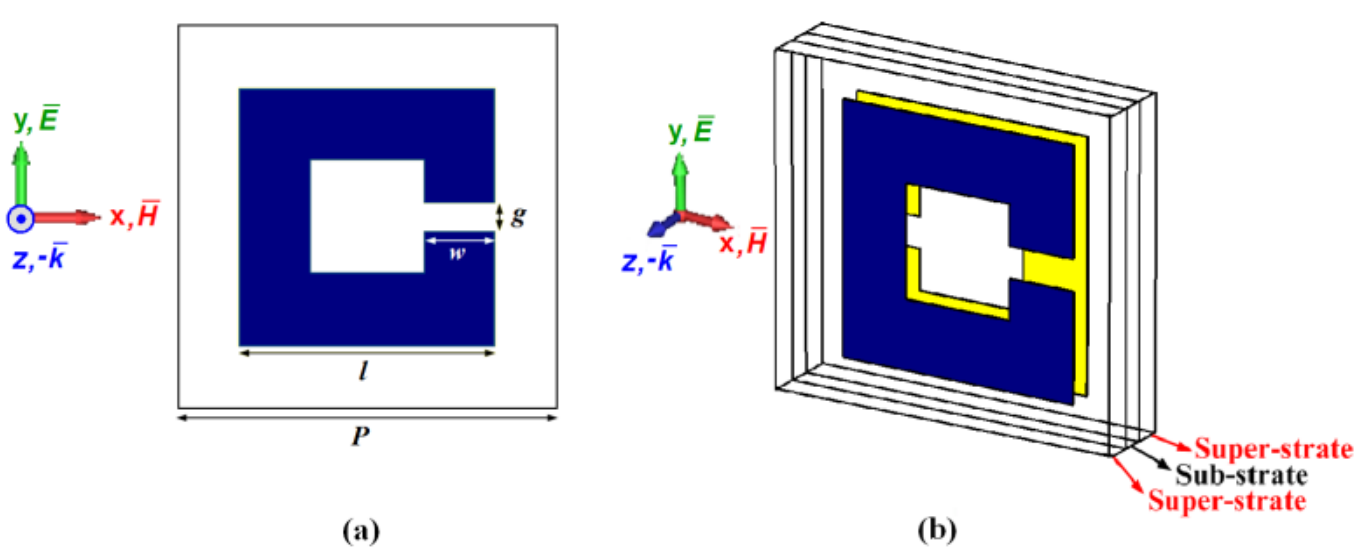

Fig. 3. Schematic view, dimensions, and excitation for a single SRR, which forms BC-SRR structure. *P*=58µm, *l*=40µm, *w*=11µm, *g*=5µm, substrate thickness ($t_{sub}$)=5 µm, and super-strate thickness ($t_{super}$)=5 µm. (a) Top view. (b) Transparent perspective view for the BC-SRR case with no-shift.

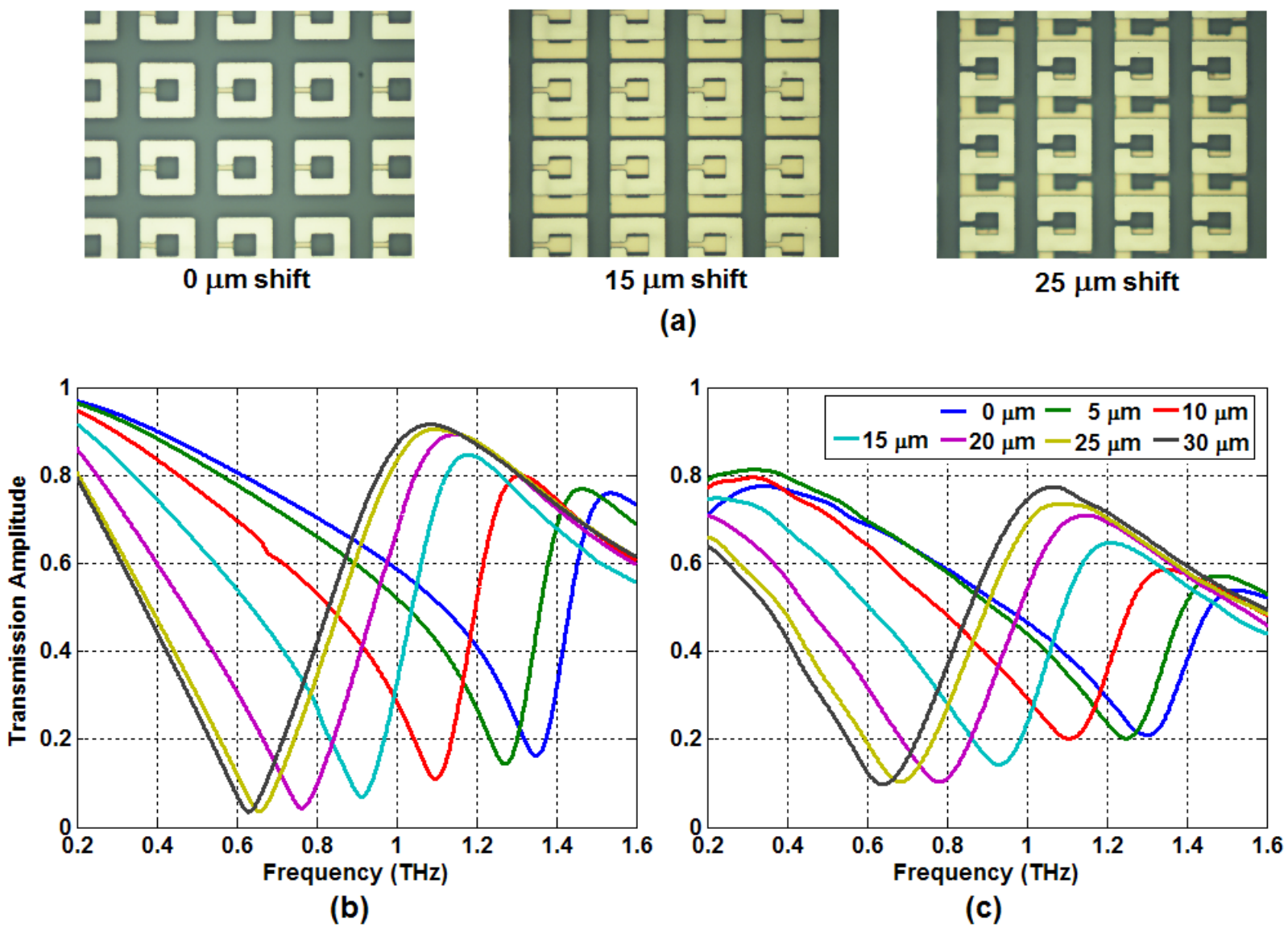

Fig. 4. (a) Optical microscope pictures of some specific BC-SRR structures shifted along vertical direction for 0 µm (no-shift), 15 µm, and 25µm. (b) Simulation and (c) experimental results for transmission characteristics of BC-SRR structures shifted along vertical direction for 0µm, 5µm, 10µm, 15µm, 20µm, 25µm, and 30µm.

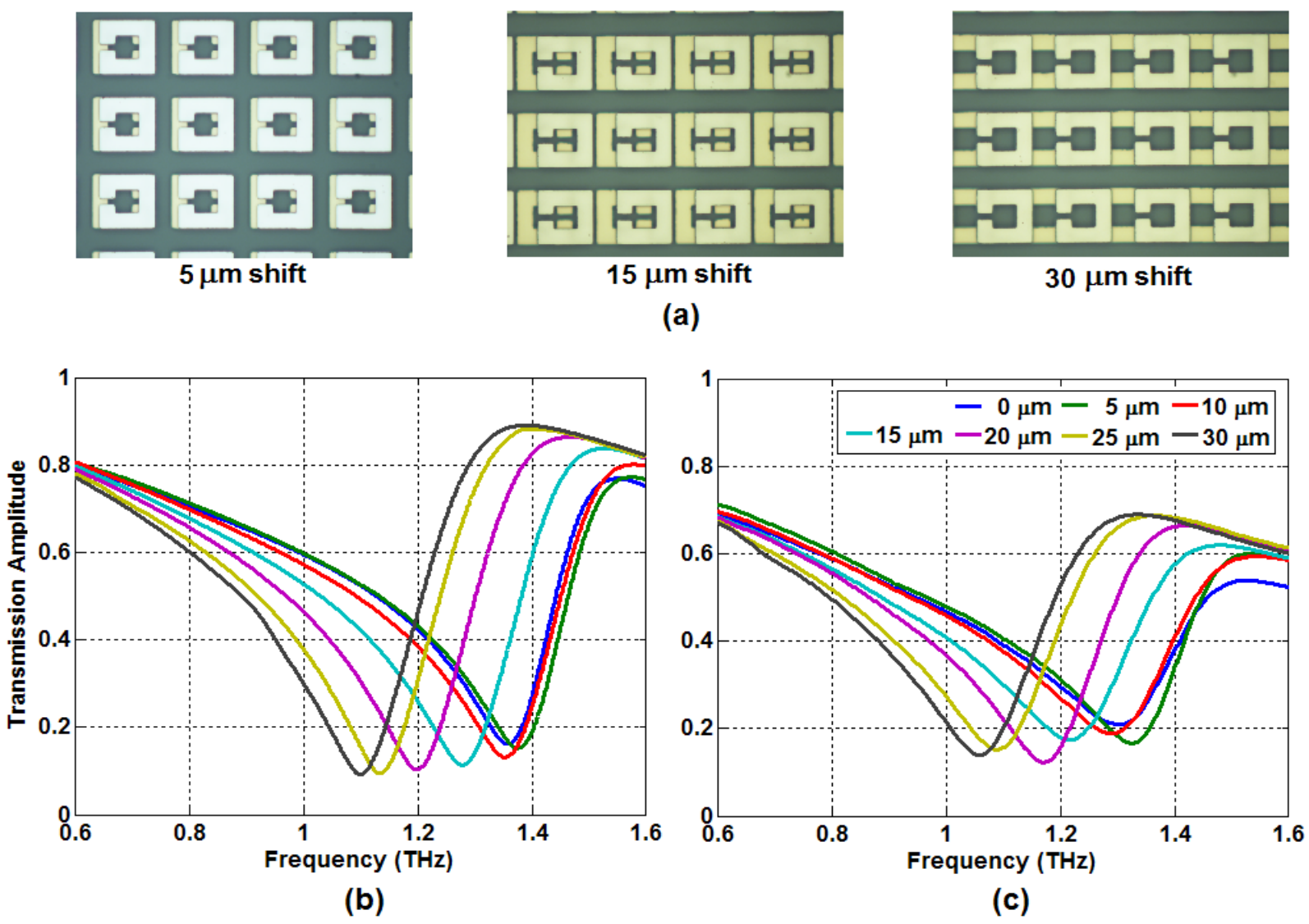

Fig. 5. (a) Optical microscope pictures of some specific BC-SRR structures shifted along horizontal direction for 0µm, 15µm, and 30µm. (b) Simulation and (c) experimental results for transmission characteristics of BC-SRR structures shifted along horizontal direction for 0µm, 5µm, 10µm, 15µm, 20µm, 25µm, and 30µm.